\documentclass[sigconf, table]{acmart}
\usepackage{xcolor}
\usepackage{fdsymbol}
\usepackage{xspace}
\usepackage{tabularx}
\AtBeginDocument{%
  }

\setcopyright{none}
\acmDOI{}
\acmConference[LIMITS '25]{11th Workshop on Computing within Limits}{June 26--27, 2025}{Online}
\acmISBN{}

\begin{document}

\title{Zombitron: towards a toolbox for repurposing obsolete smartphones into new interactive systems}

\author{Clara Rigaud}
\email{allo@clararigaud.com}
\orcid{0000-0002-6673-0546}

\renewcommand{\shortauthors}{Clara Rigaud}
\newcommand{\etal}{et~al.}
\newcommand{\todo}[1]{\textcolor{red}{TODO: #1}}
\newcommand{\yes}{$\medblacksquare$\@\xspace}
\newcommand{\yesb}{$\medsquare$\@\xspace}
\newcommand{\node}{$\textit{Node.js}$\@\xspace}
\newcommand{\termux}{$\textit{Termux}$\@\xspace}
\newcommand{\zombitron}{$\textit{Zombitron}$\@\xspace}
\newcommand{\n}[1]{$(\texttt{#1})$\@\xspace}
\newcommand{\jean}{Jean Turner\@\xspace}
\newcommand{\suf}{s8jfou\@\xspace}

\begin{abstract}
This article explores the possibilities of reusing obsolete smartphones and tablets to build new interactive systems. Taking the case of a musical instrument, I present my research into the design of a controller made from various of these obsolete smartphones.
From the diagnostic stage to the creation of a new autonomous electronic object, I document the process, the barriers and the levers encountered.
Based on these explorations and discussions with two professional musicians, I provide several insights into the software and hardware aspects, with a view to continuing this work, towards the creation of an open-source toolkit enabling anyone to build new interactive systems with old devices. I discuss the implication of how a high-level web-based approach could allow designers to enter the black box and foster permacomputing using smartphones.
\end{abstract}





\maketitle

\section{Introduction}

As electronic devices become increasingly widespread across the globe, mountains of electronic waste are reaching new heights.
A recent report of the Global E-waste Monitor \cite{balde_global_2024} estimates that 62 billion kg of electronic waste has been generated worldwide in 2022.
Of this 62 billion, only 22\% is documented as being recycled in a controlled and environmentally-friendly way and 22\% is stored as waste in open dumps.
The remaining 56\% is recycled in uncontrolled circuits, which may release materials such as mercury or plastic into the environment exposing people to unhealthy work and unsanitary conditions \cite{rifat_breaking_2019}.

Smartphones and other communication devices account for 8\% of this waste.
They are often abandoned because the battery no longer works, the screen is broken, they do not have enough memory space, the software is obsolete or they are replaced with more recent technology \cite{prabhu_n_disposal_2023, mosesso_obsolescence_2023}.
Most of the time, these devices are kept by their owners and are not recycled, repaired or reused \cite{prabhu_n_disposal_2023}.
This means that there are plenty of phones lying dormant in the drawers of millions of people \cite{ademe_longue_2023, wilson_hibernating_2017}.

Of course, work is being done on how to reduce this waste generated by smartphones. In particular, the main approach is to extend their lifespan, which is not always easy, given that many manufacturers make their equipment obsolete, and even base their business models on this obsolescence. More recently, legislative constraints have been implemented, and initiatives such as the reparability index or durability index have appeared in France \cite{ademe_preparatory_2021,noauthor_decret_2024,noauthor_decret_2020} to limit this obsolescence, but this only applies to the most recently built devices, leaving a whole range of these smartphones out of use. Another way of extending the life of smartphones is to refurbish them and buy them second-hand when the owner wants to get rid of them, but this only concerns the most recent models as well, leaving obsolete models aside.

The case I am interested in is the one where technologies are still functional but have been rendered obsolete anyway because they cannot keep up with current usage. For example, because they only have 3G, they're no longer powerful enough to support the latest software updates, or their storage memory is saturated  \cite{mosesso_obsolescence_2023,magnier_replaced_2022}.
Although there are ways of prolonging the use of one's smartphone, 
prolonging the life of an obsolete device requires compromises in terms of use or advanced technical knowledge. 

In this article, I propose a new approach consisting of combining several obsolete smartphones with each other to enable the creation of a new interactive system. I explore the possibilities that old devices can offer with an approach as high-level as possible, that does not involve reinstalling a custom OS, and that does not require a high degree of coding knowledge.
After a brief overview of the existing works, I present the \zombitron project, which consists of setting up a toolbox of design and open-source software, and detail my explorations in the construction of this toolbox.
Then, on the basis of these initial explorations, I document my research into the design of a new interactive system from the very first stage of getting it up and running. Using 14 discarded smartphones, I detail step-by-step the operations and problems encountered, and how these can be work-around, depending on the version of the operating system (OS), their performance and their general state of repair.
Then, I present the discussions I had with two professional musicians on how this tool could be integrated into their work and practice, giving promising leads as to the appropriation of this approach.
Finally, I explore the possible directions this work could take, in particular in the context of musical interfaces, to enable rapid prototyping, or as a pedagogical tool.
\section{Background and related work}
Smartphones and other communication devices account for 8\% of the total of E-waste generated worldwide \cite{balde_global_2024}. Obsolescence is the main reason why connected devices end-up being discarded, often because of limitations in terms of storage memory, computing capacity or communication protocols, which no longer allow the device to function in a wider technical ecosystem \cite{boano_enabling_2021}.
%

In addition to obsolescence, smartphones and tablets, compared to other communication devices, are also affected by a phenomenon of frequent renewal linked to the evolution of the perception of the value of their device by their owner \cite{magnier_replaced_2022}. The average length of use of smartphones by their first owner is estimated at 3.5 years in Western Europe \cite{magnier_replaced_2022}.
Smartphone obsolescence is particularly linked to rapidly evolving technology and therefore to software that becomes more demanding in terms of resources and is only maintained a few years after its release.
For example, in the case of Apple products, the latest OS version supported at the time of writing is IOS 18~\footnote{\url{https://endoflife.date/ios}}, and the last device to be able to install version 17 is the iPhone XS\footnote{\url{https://en.wikipedia.org/wiki/IOS_version_history\#Hardware_support}} released in 2018. This means that all Apple smartphones released before 2018 can no longer get updates (including security updates). 
Android devices suffer from the same problem, with the difference that it is possible to install more recent versions of Android on older devices, which are more resources demanding and may not be suited to the configuration of an older phone.
Generally speaking, the applications that are loaded onto smartphones and tablets are becoming more and more memory- and compute-intensive, slowing down devices and limiting their overall usability \cite{mosesso_obsolescence_2023}. 

The main problem with smartphones renewal, is knowing what to do with them once they are no longer in use. The first approach is to reintegrate them into recovery circuits, either by recycling them, which recovers the rare metals that make up these devices (and therefore reduces the devastating extraction of these components), or by extending their life through refurbishing services or reuse strategies \cite{prabhu_n_disposal_2023}. Refurbishing smartphones is a quite popular strategy since it has advantages for both manufacturers and consumers, but it is only reserved for the most recent products, and does not guarantee the lifespan or reliability of the refurbished device \cite{hazelwood_life_2021}. Furthermore, the recycling and reuse circuits are not always beneficial from an environmental and health point of view due to the infrastructures and standards in different parts of the world \cite{balde_global_2024} and the rebound effects linked to energy expenditure and greenhouse gas emissions \cite{zink_comparative_2014,makov_does_2018}.
In addition, many of these devices are not integrated into the recycling or refurbishing circuits at all.  Wilson~\etal \cite{wilson_hibernating_2017} found in their 2017 study that over 50\% of participants keep at least one mobile phone at home that they have replaced, usually ``as a spare''.
In France, ADEME \cite{ademe_longue_2023} estimates that 1 million smartphones are lying dormant in French people's drawers.

Studies have explored how to extend the life of obsolete devices, through alternatives and strategies for continuing to use them as their main smartphones. For example, Mosesso \etal \cite{mosesso_obsolescence_2023} have studied the different reasons and strategies used by people who continue to use their obsolete smartphones. Goodwin \etal \cite{goodwin_quantifying_2023} explored the availability of applications on obsolete Apple devices. 
These two studies conclude that it is possible to find ways of continuing to use ageing smartphones, but that these strategies often require compromises or extra effort that not all users are prepared to commit to. Some of these techniques may as well involve extra skills, for example to install new alternative OSes\footnote{\url{https://en.wikipedia.org/wiki/List_of_custom_Android_distributions}} on Android smartphones.

Another approach to reuse is to repurpose obsolete smartphones \cite{hansson_decade_2021, zink_comparative_2014, blevis_sustainable_2007} so that they can be specialised to a precise task, better suited to their performance.
For example Switzer~\etal \cite{switzer_junkyard_2023} explored techniques to repurpose smartphones into specialised servers for cloud computing. Klugman~\etal \cite{klugman_android_2018} and Norbisrath~\etal \cite{norbisrath_empowering_2025} explored how to turn them into gateways for IOT.
Smartphones and tablets can also be repurposed to be dedicated to a particular application downloaded into it, for example in educational contexts \cite{li_smartphone_2010}. 
The main obstacles in this repurposing lie in the heterogeneity of devices \cite{li_smartphone_2010} and the relatively expert operations involved, if this repurposing goes beyond simply downloading a dedicated application. In addition, Apple smartphones are poorly represented in smartphone repurposing projects as they do not allow the installation of third party applications. 

Repurposing techniques are popular in the Do It Yourself and hacking communities \cite{wakkary_sustainable_2013,lu_unmaking_2024}, and are attracting widespread interest from researchers in sustainable human computer interaction \cite{jackson_breakdown_2014}.
In particular, the burgeoning permacomputing \cite{mansoux_permacomputing_2023} movement is interested in what can be done with what exists in terms of technological hardware, and in particular in creation. This movement brings together not so recent approaches of creation but which, now, in the current context, are tinged with a particular urgency. Reuse in design has been a particular focus of Nicolas Nova \cite{nova_du_2020}, who presents the reuse of old technologies as a low-tech approach to design. Some new media artists are incorporating the reuse of dysfunctional and obsolete objects into their artistic practice, in line with the Zombie-media concept \cite{hertz_zombie_2012}. This is the case, for example, of Recyclism\footnote{\url{https://www.recyclism.com/}}, who have been exploring the revival of obsolete technologies for many years now. Others use electronic waste in their work to send out a message about technological consumption \cite{argabrite_technology_2022}. The permacomputing approach is also closely linked to that of demoscene, which since the 80s has been exploring the most restrictive ways of creating animations or music on the few remaining bits available on the first game consoles, and still today continues to gather around ``release parties'' where everyone comes to take part in demo\footnote{\url{https://www.pouet.net/prodlist.php}} competitions in their own category (weight/platform).

While these DIY communities are reputed for their openness and habit of sharing knowledge and technique \cite{wakkary_sustainable_2013,rigaud_ressources_2023}, the fact remains that these practices are reserved for relatively expert users.
Electronic objects have become smaller and complex \cite{girard_computing_2024}, and the smartphone has become a little box filled with different sensors that are difficult to reuse and tamper with \cite{nova_figures_2018}. A few soldering points are no longer enough to hack and bend must devices, leaving the reuse of obsoletes smartphones relatively absent from these projects.

In response to the increasing complexity of electronic devices, which makes certain sustainable practices difficult for the general public to access, Roedl~\etal \cite{roedl_sustainable_2015} encourages HCI researchers to develop tools to help with the appropriation of these practices and techniques.
Similarly, Lu \etal \cite{lu_unmaking_2024} argues for a field of HCI research focusing on the design of tools for recycling, recovery and reuse. 

In a nutshell, there are initiatives and contexts in which the reuse of obsolete smartphones can be explored. But the complexity of these devices and their heterogeneity require advanced technical knowledge, which can act as a barrier to their reuse.
It is therefore crucial to design tools to encourage and facilitate the appropriation of techniques for repurposing these obsolete devices, and fully integrate obsolete smartphones in the permacomputing garden.
\section{Zombitron project and principle}
The aim of this article is to present the initial work on the design of new interactive systems using a high-level approach through the \zombitron project.
The aim of \zombitron is to enable new systems to be designed from obsolete smartphones and tablets, taking into account their heterogeneity.
Given that these devices offer many possibilities in terms of connectivity and interactivity, the objective is to explore how we can continue to benefit from them.
\zombitron aims to make the design of new systems based on these devices accessible by facilitating access to the functionalities of the black box that are smartphones and tablets, and to become an open-source and collaborative tool enabling artists and everyone to address these issues of obsolescence in the case of the smartphone.

\subsection{Principle}
\begin{figure}[ht]
    \includegraphics[width=\linewidth]{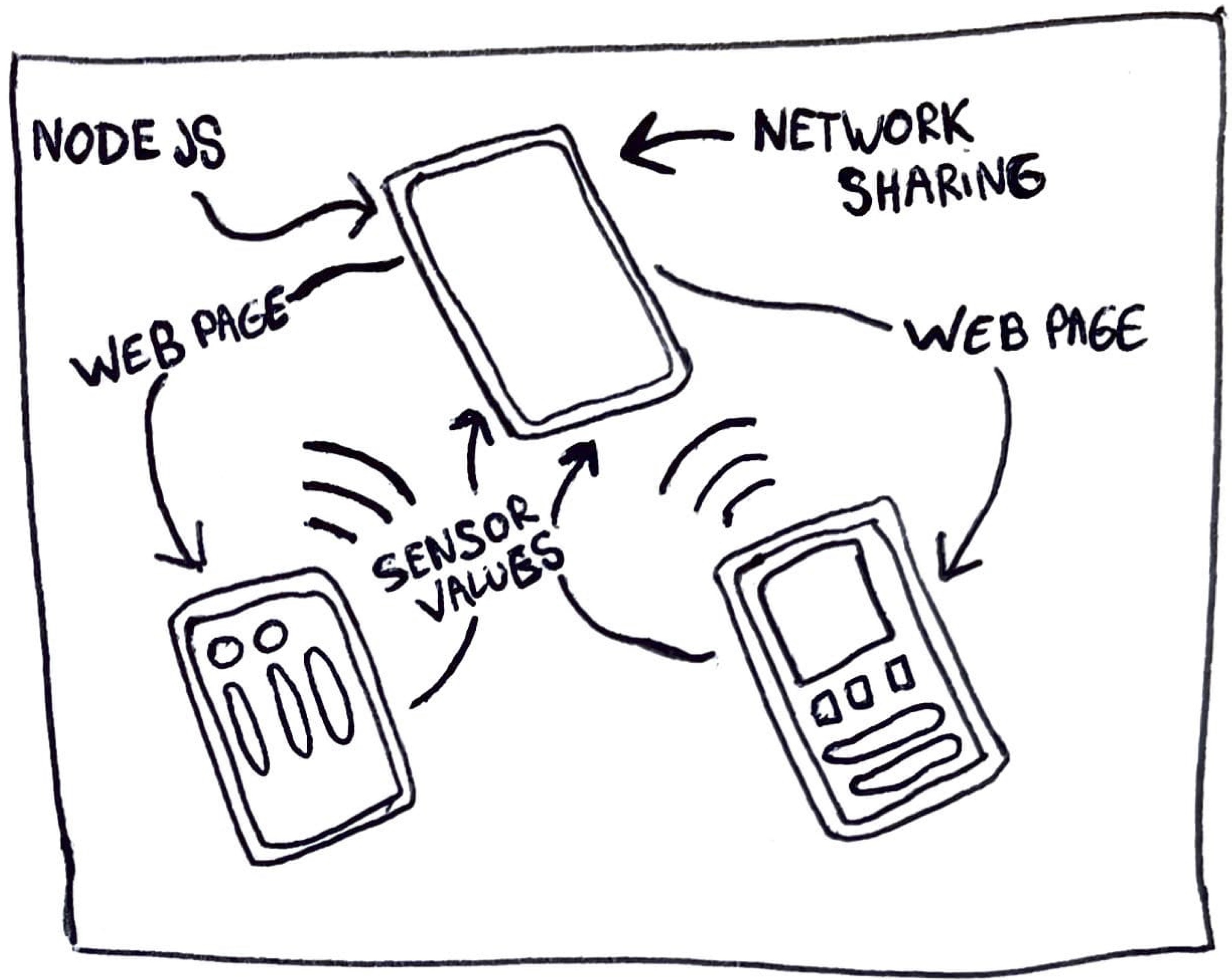}
    \caption{The basic idea behind \zombitron is to connect several smartphones together to share their information.}
    \label{fig:zombitronprinciple}
\end{figure}
The \zombitron principle is based on the fact that whatever the release date of a smartphone and its OS, it has a WiFi card and a browser. From there, it is possible to connect these smartphones and tablets to a server that displays web interfaces and exchanges information via the WiFi connection.

As shown (Figure~\ref{fig:zombitronprinciple}), the system is composed of different units that can be any connected device with a browser: 
\begin{enumerate}
    \item A unit that acts as a hotspot so that others can connect to it locally.
    \item A unit that runs a terminal enabling a \node instance to be run and an \texttt{http} server to be generated on this network, also enabling the \texttt{websocket} protocol. 
    \item One or more units that will connect to the server as web clients.
\end{enumerate}

For example, depending on its configuration, a unit can act as a hotspot (1), run the server (2) and display a web interface (3). It is also possible to use a computer to run the server and an external router to create a local network.

The way in which the system is implemented will essentially depend on the capabilities of each of the units: 

\subsubsection{Creating a local network}
There are several ways of creating a local network. The simplest way is to use a dedicated router, for example an internet box. But in the case of an interface designed to be autonomous and operate without any devices other than the smartphones and tablets it is made up of, a shared connection needs to be set up.
To do this, it is possible to activate tethering mode on certain devices. This option is widely available, but some OSes do not allow this option to be activated if there is no access to mobile data.
The unit bundle must therefore include at least one device enabling tethering without mobile data. 

\subsubsection{Running a server with \node}
To run a \node server, one of the units needs to be able to run a terminal. To do this, \zombitron uses the \termux application, which emulates a Linux terminal on Android devices.
This enables packages to be downloaded using ``apt'' and \node to be run.
With \node, an \texttt{http} server is set up to deliver webpages to other units connected to the network, and a \texttt{websockets} server is used to send data between units in real time. 

\subsubsection{Displaying web interfaces}
\begin{figure}[ht]
    \includegraphics[width=\linewidth]{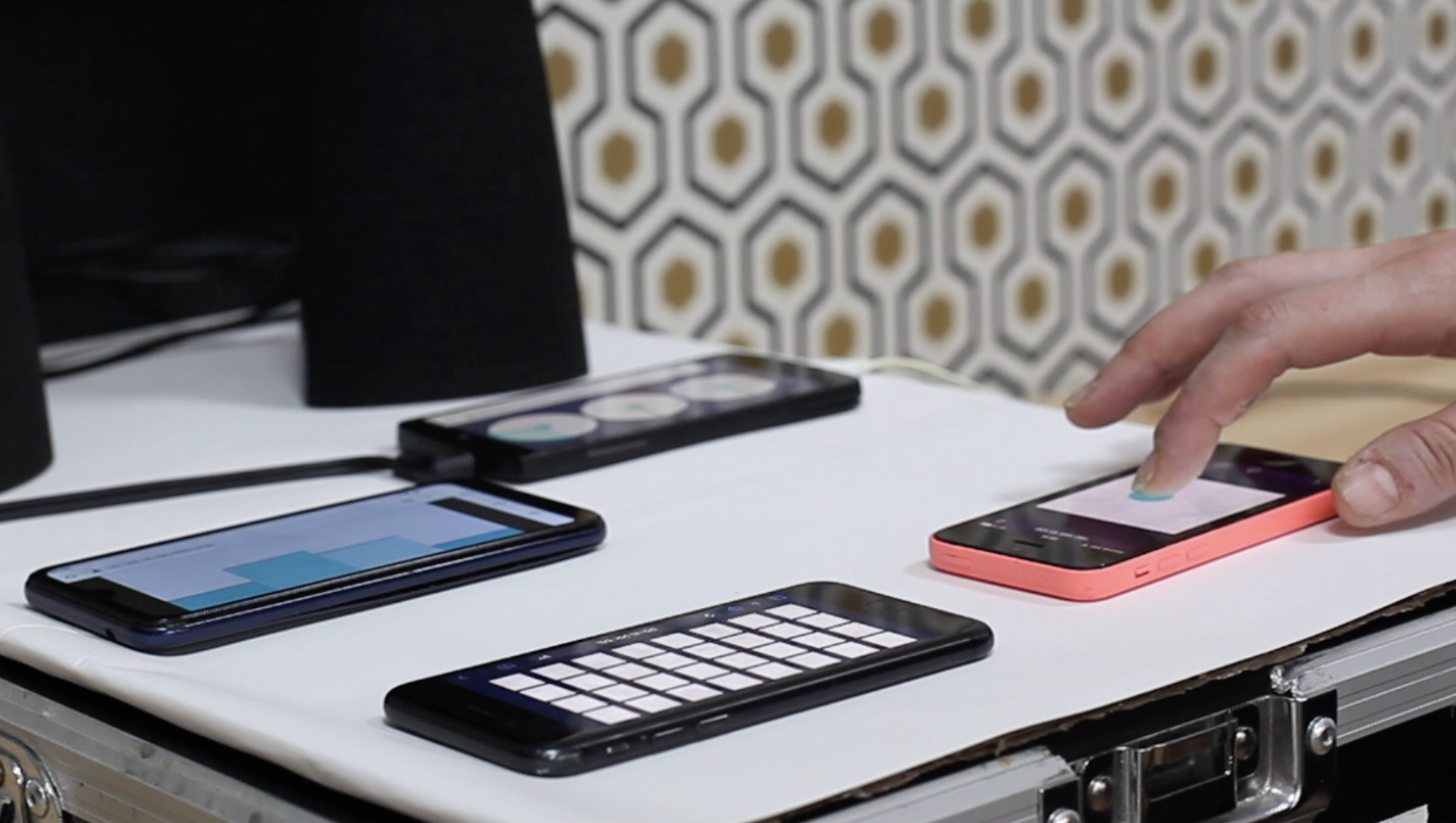}
    \caption{This images shows how one can create a new interface by loading webpages on different smartphones.}
    \label{fig:principlewebpage}
\end{figure}

Once this environment has been set up, it is possible to set up any web application that is capable of being executed on the smartphone browser. For example, figure \ref{fig:principlewebpage} shows the control interfaces loaded on different smartphones composing music. 
Each smartphone displays a webpage and can send messages related to touch events to the other units. 

\subsection{First explorations}
Based on this principle, two musical instrument prototypes have been designed: \textit{Zombitronica} and \textit{Zombichord}.
These two prototypes are built using the \textit{Tone.js} library dedicated to sound synthesis through web audio \cite{jensenius_2013_2017}.
From a hardware point of view, both are built from reused materials, and operate in standalone mode.

\subsubsection{Zombitronica}
\begin{figure}[ht]
    \includegraphics[width=\linewidth]{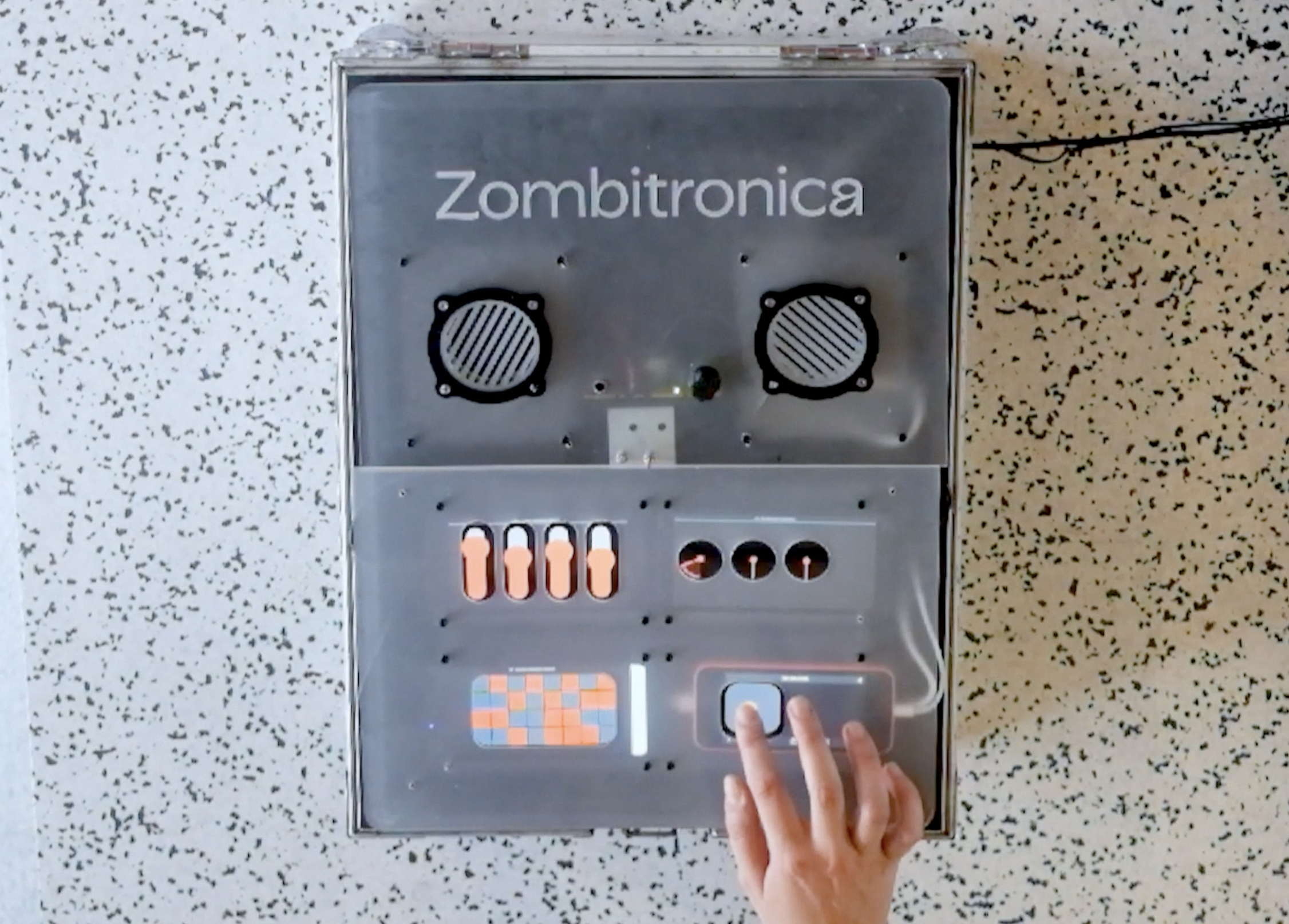}
    \caption{Zombitronica is the first prototype exploring the principle of \zombitron, made with 4 discarded smartphones and a loud-speaker.}
    \label{fig:zombitronicacontrollers}
\end{figure}
Zombitronica is the first prototype based on 4 smartphones and a loudspeaker, all second-hand.
The interface has been designed to explore different types of touch controllers for generating music (Figure \ref{fig:zombitronicacontrollers}). 
It is made up of a sequencer for selecting the notes played by the 4 instruments over 8 beats, 4 sliders for controlling the volume of each of the instruments, a two-dimensional slider for controlling a 5th instrument (its pitch and volume) and 3 potentiometers controlling the speed of the metronome, distortion and reverb.
From a hardware point of view, all the phones are powered by a usb hub, and the phone generating the music is connected to the speaker via a 3,5mm jack.
A laser-cut sheet of 3mm PMMA delimits the screen areas corresponding to the control interfaces, and holds the 4 smartphones and the loudspeaker in place. Each component rests on a plate suspended from the top by a system of screws, nuts and spacers. 

\subsubsection{Zombichord}
\begin{figure}[ht]
    \includegraphics[width=\linewidth]{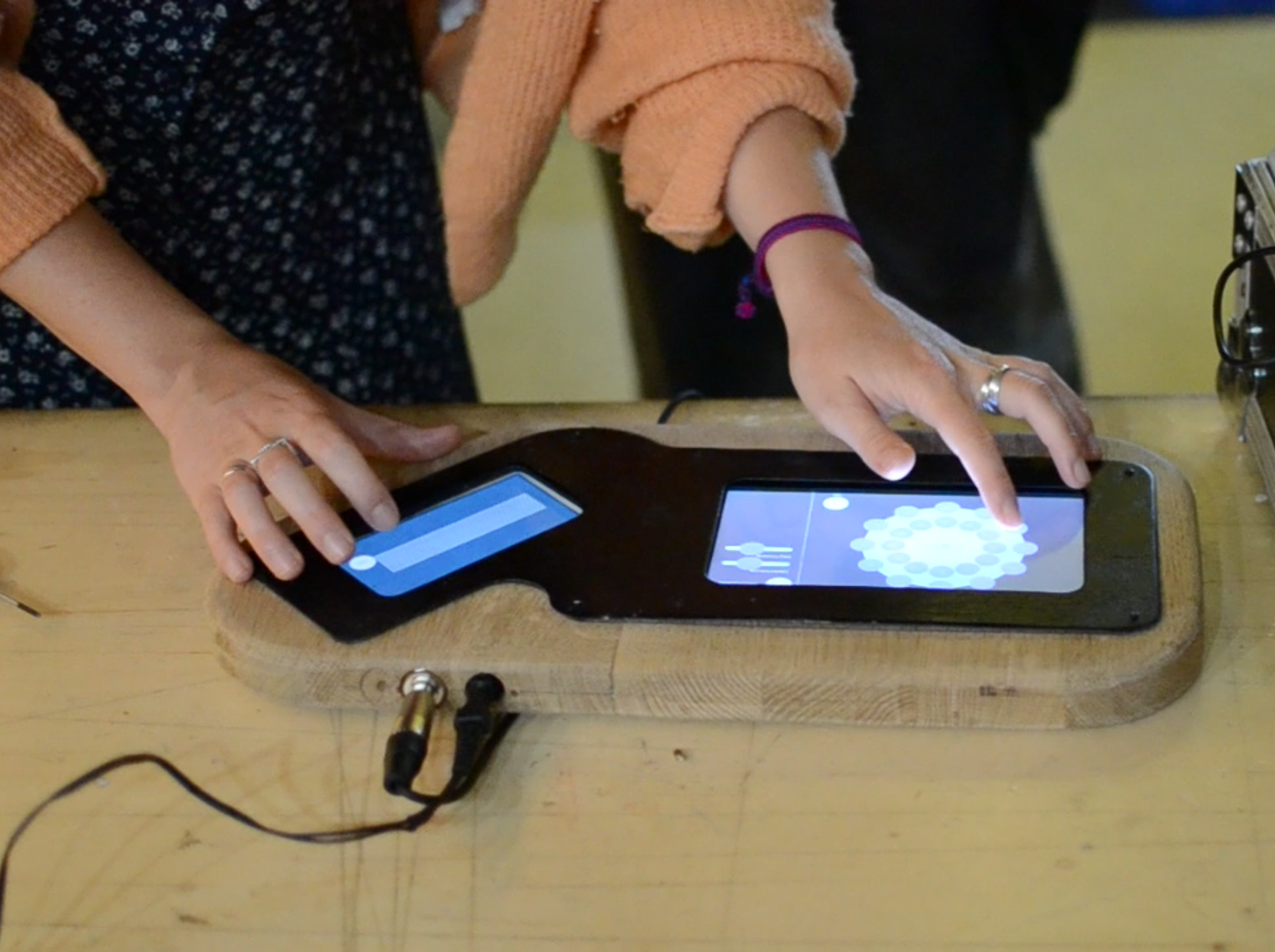}
    \caption{\textit{Zombichord} is a the second prototype exploring the principle of \zombitron. It was designed by Romain Segaud and Marion Jolas, inspired by the Omnichord.}
    \label{fig:zombichordinterface}
\end{figure}

\textit{Zombichord} is a second prototype inspired by the Omnichord\footnote{\url{https://en.wikipedia.org/wiki/Omnichord}} and built using an Android 5 tablet and an IOS smartphone.
On this prototype, the \node server is run on the tablet, which also displays an interface for playing chords, and an iPhone 6 displaying an interface for playing notes on the scale corresponding to that chord [figure \ref{fig:zombichordinterface}].
For the hardware, a thick piece of wood was hollowed out with a wood router to accommodate the two devices. Holes were drilled to allow the lightning-type cable connected to the iPhone to pass through on one side and a mini usb cable to supply power to the tablet on the other. The lightning cable is used to output both the sound signal and to supply power to the phone, so a few soldering points were used to make a small assembly that simply reveals a female jack (for the sound output) and a female coaxial socket for supplying 5V to the two devices. 

\subsection{Outcome}
These two prototypes enabled me to test a web approach to creating two musical interfaces on discarded smartphones and tablets.
Although only the touch screen was used, they already enabled me to identify certain opportunities and limitations linked to the age of these devices, particularly in terms of performance and functionality. 
The choice of implementing the software in webpage format on older smartphones, which have less computing power than today's products, may mean that we have to use lightweight applications. Devices that only load a touch interface and send their data via \texttt{websockets} did not pose any problem, but the use of a library that requires a little more CPU processing, such as \textit{Tone.js} doing sound synthesis, should be considered for rather more powerful smartphones.
In order to eliminate the latency of the two prototypes, the use of \textit{Tone.js} was restricted to the activation of pre-recorded samples. This opens up opportunities to explore how the combination of older and more efficient equipment can overcome these limitations.
The second prototype was designed using a tablet running Android 5.0. For this tablet, the \termux application was no longer supported, which means that the package database is no longer updated, limiting the version of \node to version 12. A few readjustments to the code enabled me to build the application with \node 12 without any problems.
\section{Studying the possibilities offered by obsolete smartphones}
On the basis of the two explorations made through the \textit{Zombitronica} and \textit{Zombichord} prototypes, I have sought to confront this approach with a more general situation, to consider the limits that a variety of these old devices may have, and the ways in which they can be overcome.
With a view to creating a common tool that would enable everyone to design with obsolete devices, I looked at a wider range of these devices to document, step by step, the specifics in terms of installation and configuration that I may encounter.
I collected a set of 14 smartphones that had been given to me or that I had bought cheaply second-hand. For each of the phones, I went through the different stages to set up the basis of the \zombitron principle. 
This time, in addition to the touch sensor, I integrated the Inertial Measurement Units (IMU) sensors: gyroscope and accelerometer, common to most smartphones since 2012.

During this exploration, I set up a common code base on a git repository\footnote{The code base used in this article is preserved here: \url{https://zombitron.clararigaud.com}}, allowing me to run tests on the different devices, and which I improved as I went along so that it took into account the particularities of each one.
This code can then be used as the basis for all interfaces created using the \zombitron principle, and can be expanded as we implement them. Additional resources on specific commands I have used can be found in Appendix \ref{sec:supplementary}.

In this section, I break down the steps from the first switch-on to loading the webpage and sending the sensor data, documenting the obstacles I encountered depending on the device.

\begin{figure}[h]
    \includegraphics[width=\linewidth]{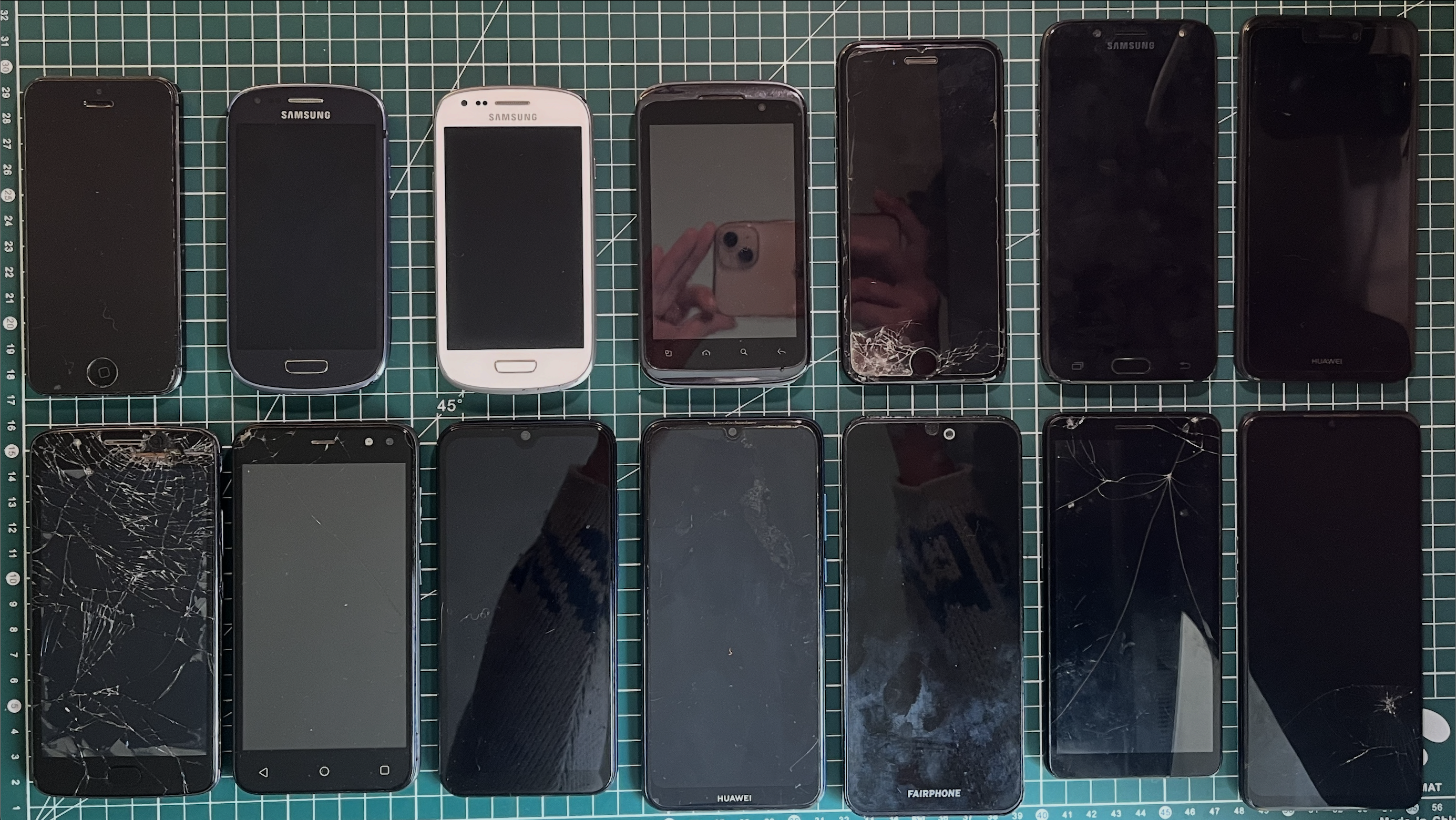}
    \caption{Overview of the 14 discarded smartphones batch.}
    \label{fig:allphones}
\end{figure}

\definecolor{LightCyan}{rgb}{1, 0.45, 0.65}

\subsection{Turning on and unlocking}
\newcolumntype{Y}{>{\centering\arraybackslash}X}
\begin{table}[ht!]
    \raggedright
    \begin{tabularx}{\linewidth}{@{}clYYYY@{}}
        & \textbf{Modele}  &  \textbf{Year} & \textbf{Server} & \textbf{Hotspot} & \textbf{Client}  
    \\
    \n{1} & iPhone 5 & 2012 & & & \yes
    \\
    \n{2} & SM Galaxy S III & 2012 &  & \yes & \yes
    \\
    \rowcolor{LightCyan}
    \n{3} & SM Galaxy S III & 2012 & & & 
    \\
    \n{4} & Moche smart A8 & 2012 & & \yes & \yes
    \\
    \n{5} & iPhone 7 & 2016 & & & \yes
    \\
    \n{6} & SM Galaxy J5 & 2017 & \yes & & \yes
    \\
    \n{7} & Huawei P8 lite & 2017 & \yes & \yes & \yes
    \\
    \rowcolor{LightCyan}
    \n{8} & Moto E4& 2017 & & & 
    \\
    \n{9} & Gigaset GS80 & 2018 & \yes & \yes & \yes
    \\
    \n{10} & Archos oxygen 57 & 2019 & \yes & \yes & \yes
    \\
    \rowcolor{LightCyan}
    \n{11} & Huawei Y6 & 2019 & & & 
    \\
    \rowcolor{LightCyan}
    \n{12} & Fairphone 3 & 2019 & & & 
    \\
    \n{13} & Alcatel 1B & 2020 & \yes & \yes & \yes
    \\
    \rowcolor{LightCyan}
    \n{14} & Oppo find x2 lite & 2020 & & & 
    \\
        \end{tabularx}
        \caption{List of the 14 smartphones selected with their year of release and their ability to be used as a server unit, a hotspot unit and a client unit. Devices that could not be turned on or unlocked are marked in pink.}
    \label{tab:step1}
    \end{table}
The first step for each smartphone is to turn it on and unlock it.
Some of the phones in my batch had not been switched on for a long time, or had been considered ``bricked'' by their users.
So for each phone, I began by charging the battery, then tried to switch it on.
Some did not make it through this stage, and so I eliminated them from my batch as they require a more advanced level of repair. Table \ref{tab:step1} summarises which of the 14 smartphones could be turned on and unlocked, 5 smartphones were eliminated at this stage. 
For example, this was the case with one of the two Samsung Galaxy S III \n{3}, the Huawei Y6 \n{11}, Moto E4 \n{8} and the Oppo Find X2 \n{14}.
\n{3}, \n{8} and \n{11} had no reaction when the power button was pressed, requiring a diagnosis beyond my scope. On the other hand, \n{14} vibrated when switched on without displaying anything, suggesting a dead screen.
To check whether the devices are detected as serial usb ports (and therefore check whether they are alive), I connected the smartphones to my computer to verify whether they were detected as a USB serial interface.
The Oppo \n{14} was indeed detected as an interface. 
An alternative rescue would therefore be to try connecting it to an external display using the ``HDMI Alt Mode''\footnote{\url{ https://www.hdmi.org/spec/typec}} protocol, but this was beyond my scope. 
Interestingly, the Fairphone \n{12} did not pass this stage either, because it was locked. After switching it on, I had to enter a password that I did not have, and I did not know the person who had given it to me. 
I naively tried to carry out a factory reset on the phone, but I came up against the Factory Reset Protection protocol, which requires signing in to the last Google account associated with the device.
There are strategies, but they are time consuming, require scrolling many forums and ``adventurous'' operations, as I read on one of them.
Some of these strategies involve exploiting security breaches that are patched over time, which means that the newer, or recently updated the phone, the less easy it is to bypass the FRP. 
So it is important to make sure that devices are reset to zero before disposing of them.

\subsection{Setting up a server}
The second step is to examine which smartphones are capable of acting as a server, which involves running \node in my case.
To install and run \node, \zombitron uses the \termux\footnote{\url{https://termux.dev/en/}} application, which is only available on Android.
As a result, the two iPhones \n{1} and \n{5} in my selection will not be able to be server units.

There are several ways to download \termux, the easiest of which is to go to the Google Play Store\footnote{\url{https://play.google.com/store/apps/details?id=com.termux}} and download the application, but this is only available for devices with Android 11 (released in 2020) or later.
The other disadvantage is that this involves associating all devices with a Google account.
For these reasons, I chose to install the application without going through the Google Play Store, but rather directly by downloading the \texttt{.apk} file.
This process requires a few additional steps: installing \textit{Android Debug Bridge} (\texttt{adb})\footnote{\url{https://developer.android.com/tools/adb}}, and enabling USB debugging mode on the device to be able to load applications on the phone. These steps are widely documented on the web.
Once this is done, it is possible to easily access the smartphone's information and install \texttt{apk} files on it.
The next step is to download the \texttt{apk} file, for example from \termux's github\footnote{\url{https://github.com/termux/termux-app}}, selecting the file corresponding to the Android version and the device's CPU ABI\footnote{Android Binary Interface: \url{https://developer.android.com/ndk/guides/abis}}.
Table \ref{tab:server} gives a summary of the different versions and CPU ABI's of the Android devices in my batch.
\begin{table}[ht!]
    \raggedright
    \begin{tabularx}{\linewidth}{@{}cYYYY@{}}
    & \textbf{Android version} & \textbf{Processor's ABI} & \textbf{Termux} & \textbf{Node.js version}
    \\
    \hline
    \n{2} & 6.0.1 & {armeabi-v7a} & \yes & 12.13.0
    \\
    \hline
    \n{4} & 2.3.6 & armeabi & &  
    \\
    \hline
    \n{6} & 9 & {armeabi-v7a} &  \yes  & 22.14.0
    \\
    \hline
    \n{7} & 8.0.0 & arm64-v8a & \yes & 22.14.0  
    \\
    \hline
    \n{9} & 8.1.0 & {armeabi-v7a} & \yes  & 22.14.0
    \\
    \hline
    \n{10} & 9 & arm64-v8a &  \yes & 22.14.0
    \\
    \hline
    \n{13} & 10 & {armeabi-v7a} & \yes & 22.14.0
    \\
    \end{tabularx}
    \caption{Summary of the Android versions, the ABI of each of the android Smartphones in the batch, the availability of \termux and, if applicable, the version of \node available.}
    \label{tab:server}
\end{table}

Once \termux is installed and running, it is possible to install \node. (The smartphone must be connected to the internet first.)
For devices running Android <7, the version of \termux no longer updates the package database, which sometimes forces the use of discarded version packages\footnote{\url{https://github.com/termux/termux-app/wiki/Termux-on-android-5-or-6}}.
The installation of \termux and \node went well on the device \n{2} running Android 6. The version of \node available in Long Term Support (LTS) is \texttt{12.13.0}. 
However, an error occurred when using the \texttt{npm} command, which allows packages to be installed in \node.
Unfortunately I did not have any other devices running Android 5 or 6 to compare, but as the tablet used in the \textit{Zombichord} prototype runs Android 5, I know it is possible to use these OS versions.
\termux is available from Android 5, which eliminates my device \n{4} running Android 2.

Once \node was installed, I used git to download my test environment.
At this stage, the folder only contains a \node project and the necessary dependencies for \zombitron: \texttt{express} to create \texttt{http} servers, and \texttt{websocket} to use the real-time socket protocol for sending messages. The aim is to check that all versions of \node can generate servers with these two components, and if there are any particularities to take into account depending on the version of \node, to ensure that these are added to my code.
As a result, and after a few syntax corrections linked to the different versions, both version \texttt{12.13.0} and the current LTS version (\texttt{22.15.0}) of \node can run the server.

\subsection{Setting up a local network}
For the interface built with \zombitron to be standalone, at least one of the devices must be able to act as a hotspot to generate a local network among all the devices. 
While most devices allow this connection sharing option, some OSes require mobile data to activate this network sharing. This is the case for all IOS devices, and some Android-based interfaces like in my case here with device \n{6} which has a One UI 1.1 interface. 
Table \ref{tab:step1} summarises the devices in my sample that can be used as hotspots.

\subsection{Setting up web clients}
To test the possibilities of using each of the smartphones as a client, I developed, as in the case of the server, a code example which I put to the test on each of the devices and their browsers, and I improved it according to the problems encountered. 
I managed to get a touch interface working and to obtain IMU data from all the phones and send it to the server via \texttt{websocket}.
I detail below the obstacles encountered and how I overcame them.
\subsubsection*{Javascript ES6}
To communicate with the server, clients use the javascript language. In 2015, a major version of javascript (ES6) was released, making some older browsers incompatible.
This is the case, for example, with the browsers of \n{4}.
All the interfaces developed on \zombitron are compatible with the older version of JavaScript (ES5).

\subsubsection*{Motion sensors}
Most smartphones have two inertia sensors: the accelerometer and the gyroscope.
In my batch, all the devices have an accelerometer, and all except \n{7} \n{9} \n{10} \n{13} a gyroscope. 
These two sensors respectively provide motion and orientation data via the \texttt{DeviceOrientationEvent} and \texttt{DeviceMotionEvent} interfaces.

I was able to obtain sensor data from all the devices, taking into account the following small details in my code.

-- Safari on IOS 14.5 introduced a new security feature for access to inertial sensors: \texttt{requestDevicePermission}.

-- Access to sensor data requires a secure server (\texttt{https}) on most recent browsers, which I have enabled as an option in my code.

\subsubsection*{Secured server (https)}
A secure server is set up by creating a key and a certificate. Each certificate is linked to the server's IP address. 
As secure servers are necessary for the use of inertial sensor data, I have created a script allowing certificates to be generated directly from a command line.

Very old versions of Android Browser before 4.4.4 are incompatible with the \texttt{https} protocol. To avoid this problem on \n{4}, I installed a version of Firefox (Firefox 31) that supports \texttt{https}. The implementation of secure servers also has the effect of displaying alert pages when the interface is loaded, but these can be bypassed, often by clicking on a link such as ``I trust this website''.
\subsubsection*{Secured Websockets (wss)}
When the server is set for https, the \texttt{websocket} protocol must also be secured. This is taken into account in my code, but in the case of IOS devices < 13, it is necessary to install the certificates to use the \texttt{wss} protocol. 
To do this, the certificate needs to be downloaded, installed, then activated in the IOS settings : \texttt{Settings > General > About > Certificate Trust Settings}

\subsection{Outcome}
By taking a heterogeneous batch of smartphones, varying in their OS, release date and browser, I was able to explore in greater breadth the possibilities offered by a web approach such as \zombitron. 
This methodical and iterative work has also enabled me to develop a common code base that can run on all these devices, and that supports most of their specific features. 
This provides a base from which to create new applications and which can be fed with other functionalities and APIs.

\section{Integration into two musicians' practices}
While I have investigated the potential of a web-based approach to creating controllers from obsolete smartphones, from a technical point of view, I am now trying to understand how this tool could be integrated into real practices, using the example of music interfaces.
During my study, I had many exchanges with two friends, both professional musicians with a background in electronic music composition and performance using numerous controllers and synthesizers.
We exchanged ideas about their practices and habits in using tools to produce and perform. Then we imagined how \zombitron could enable them to build their instrument or their musical controller, and more generally how they envision the directions this tool could take.

Despite their very different practices and styles, \jean\footnote{\url{https://abstraction-mathematique.bandcamp.com/}} and \suf\footnote{\url{https://s8jfou.bandcamp.com/}} both have a strong focus on tools and machines, which they like to make their own by exploring the possibilities of customisation. Both have a particular affection for the restrictions that machines bring to their work, and at the same time the freedom that computers bring to sound design. 

\jean's musical performances are somewhere between dance and experimental music, often using samplers and sequencers that he interfaces with his computer via \textit{Ableton}\footnote{\url{https://www.ableton.com/}} and MIDI\footnote{\url{https://en.wikipedia.org/wiki/MIDI}}. On stage, his gestures and movements are at the heart of his performance. \jean has built instruments and controllers using ``Axoloti core'' \cite{charles_using_nodate} which he describes as a sort of audio and MIDI Swiss army knife, very simplified for the user. It allows him to do precise things according to his needs and to have certain functions that do not exist in other controllers.

Questioning the tools and the visual composition on stage is particularly at the heart of the work of \suf. \suf makes electronic music essentially with his computer and builds his instruments with MAX/MSP\footnote{\url{https://cycling74.com/products/max}} with which he interacts on stage. He has also built a few synthesizers\footnote{\url{https://www.s8jfou.com/synth.html}} himself using micro-controllers such as \textit{Teensy} \cite{edstrom_arduino_2016} and \textit{Raspberry PI pico} \cite{halfacree_get_2021}.

\subsection{\zombitron as a performing tool for \jean}
\begin{figure}[h!]
    \includegraphics[width=\linewidth]{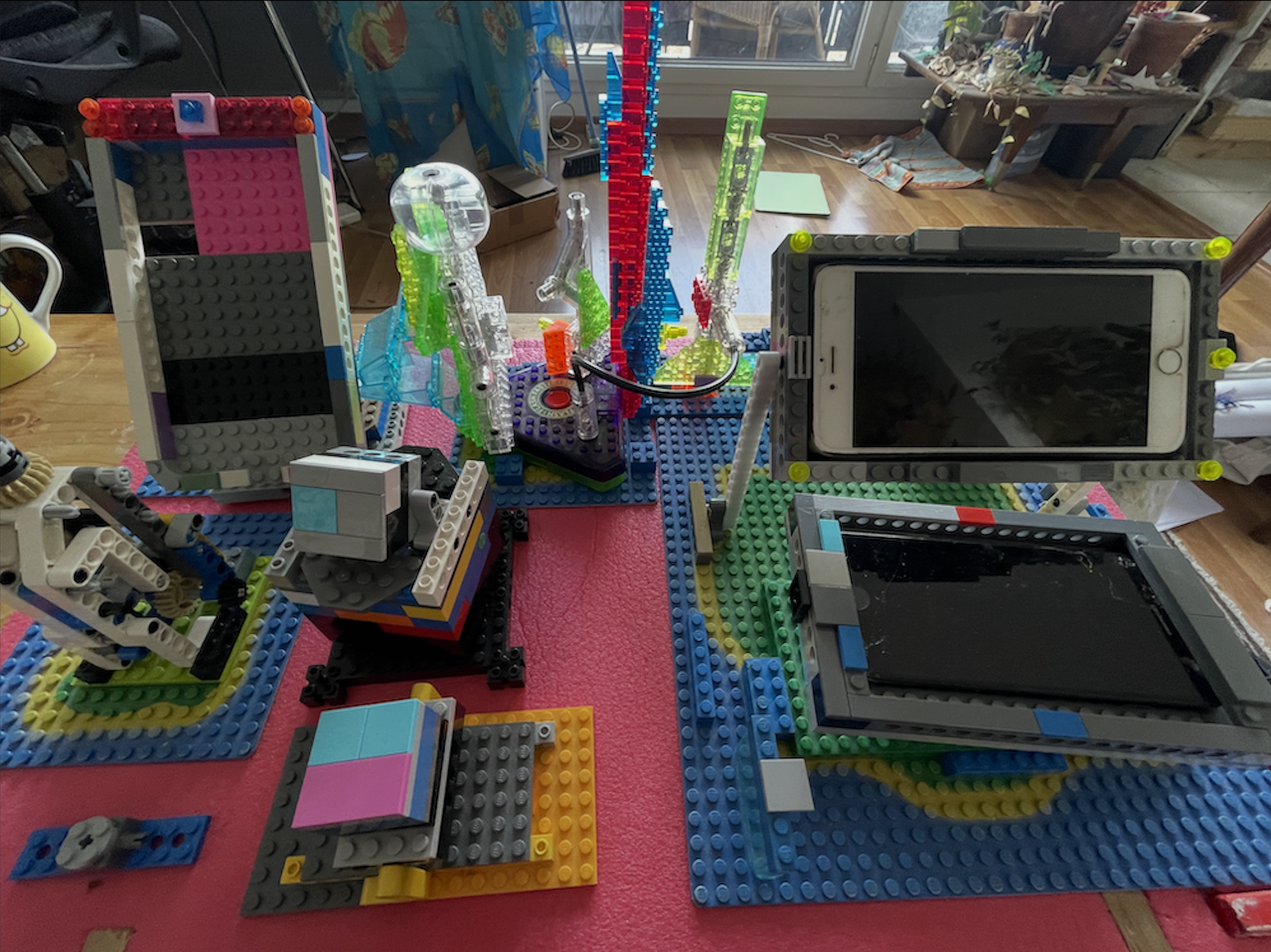}
    \caption{\jean has created different mechanisms to design an interactive interface for playing with smartphones connected to his software via Open Sound Control.}
    \label{fig:jeanturner}
\end{figure}
When I presented my work on \zombitron to \jean, he immediately imagined how he could integrate motion sensors into his scenic performance and composition.
Using smartphones to perform opens up new horizons for him. 
``\textit{I did not want to play with a computer on stage because I did not like the aesthetics. With \zombitron, it is logical because it democratises computer music even more and there is a playability that does not exist with other instruments.}''
\jean also mentions the educational aspect that this can engender in his stage performance: 
``\textit{For the audience, turning knobs does not mean much, whereas seeing the sound change with the tilt of an old smartphone is more intuitive.}''

\jean started building a whole range of articulated supports using LEGO® Technic to use smartphones as controllers connected to his Ableton software with Open Sound Control.
As shown Figure \ref{fig:jeanturner}, he created several supports to hold his two smartphones and then mounted these supports on different mechanisms, allowing them to rotate on two or three axes.
Some of these mechanisms make use of motors to activate a continuous rotation of one smartphone and lights activated by the weight of the smartphone when put in certain supports.
While he was making his LEGO®, he sent me a number of videos, in which he describes his enthusiasm as he explores the interactions, the effects he will be able to generate and the way he will compose his live show.
``\textit{I can put one on top of the other, and then my values are the same for both!}''.
``\textit{I just thought I can even use it as percussions.}''

\subsection{\texttt{Zombee}, a collaborative musical instrument for \suf}
\begin{figure}[h!]
    \includegraphics[width=\linewidth]{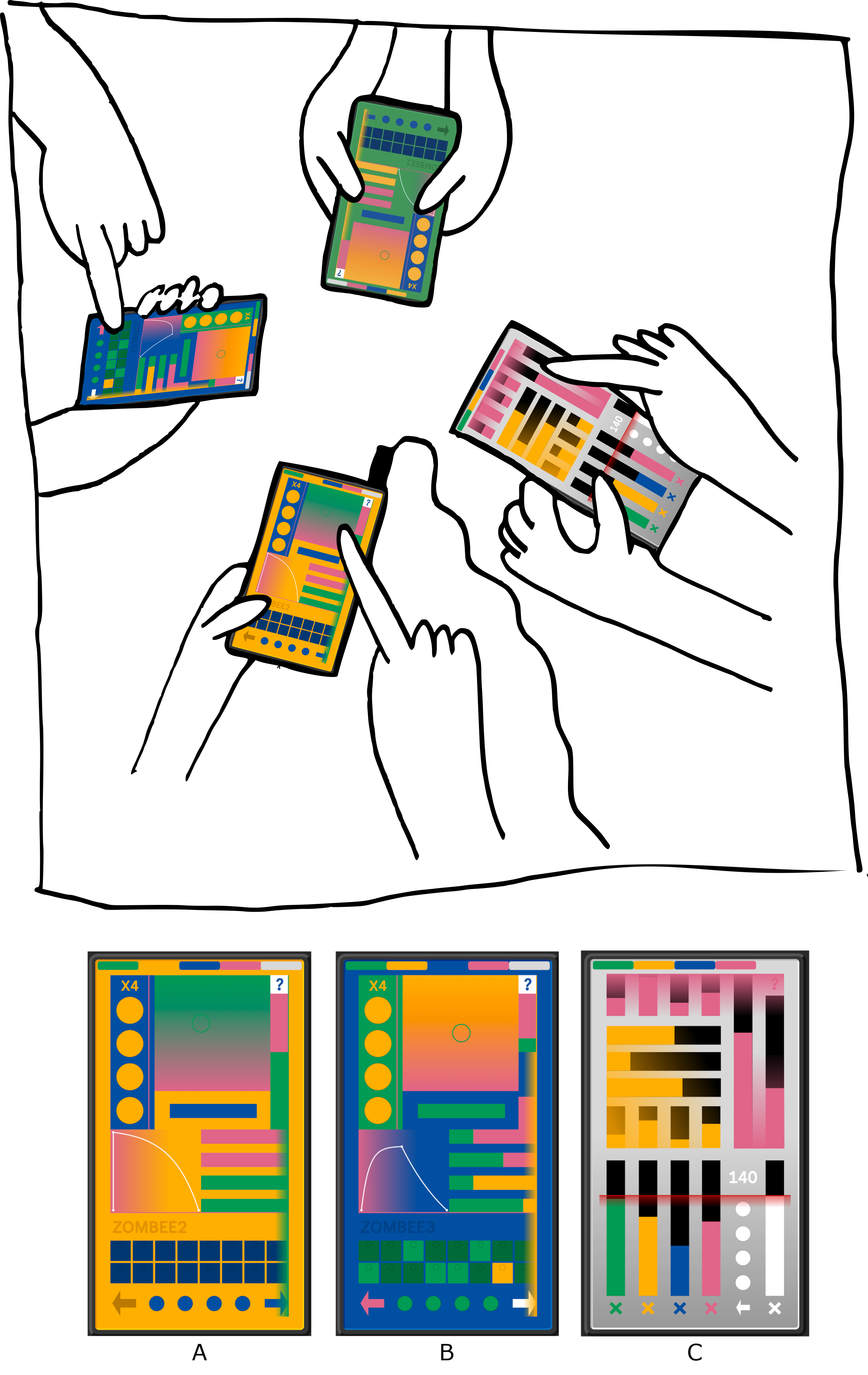}
    \caption{\suf has begun designing \texttt{Zombee}, a collaborative instrument optimised to run on devices with low processing capacity.
    A and B show two of the four composition interfaces and C shows the conductor interface.}
    \label{fig:s8jfou}
\end{figure}

\suf immediately saw in \zombitron the possibility of creating a collaborative composition environment, and was stimulated by the idea of generating sound capable of being played by devices with very limited computing power.

In 2022, \textit{MAX} released \textit{RNBO}, a library for exporting \textit{MAX} projects (patches) in portable code to compile them directly for given platforms, in particular WebAssembly. They have also released an \texttt{npm} package~\footnote{\url{https://www.npmjs.com/package/@rnbo}} that allows this compiled code to be loaded directly via \node.
From there, it was easy to imagine how to combine \suf's practice of \textit{MAX} and \zombitron.

\suf has designed \texttt{Zombee}, which he began to implement in \textit{MAX}. \texttt{Zombee} is an instrument consisting of 4 playing interfaces and a master interface, enabling several people to compose and play together.
The 4 tactile interfaces (Figure \ref{fig:s8jfou} \texttt{(A)} and \texttt{(B)}) differ simply by the sample bank, and can be percussive, bass, synth, noise, etc. To this sound ``search'' interface, he added a sequencer to be able to compose. 
Added to these 4 instruments, he created an interface dedicated to the controlling the different parameters of the 4 instances (Figure \ref{fig:s8jfou} \texttt{(C)}), the volume, the BPM, the different effects (Delay, Reverb,...). ``\textit{A bit like an orchestra conductor}''.
\suf envisaged \texttt{Zombee} by thinking about the limits in terms of memory and computing capacity that the device generating the sound might have. To achieve this, he uses the principle of granular synthesis \cite{roads_introduction_1988}, involving the use of samples of very short duration so that relatively light files can be loaded. 
He has designed his instrument so as to connect all the effects to a single output managed by the conductor, which also means that the effects (which consume a lot of CPU resources) are applied only once.
He also identifies the advantage of versatility offered by this approach, imagining how it could be both stand-alone and used with a more powerful computer.
``\textit{Depending on the power of the device used, more or less can be done in sound synthesis. The tool is therefore very powerful if you want to use it, for example, with a recent computer but with old phones to control it remotely, but is also capable of doing more modest things when run from an old phone.}''

\subsection{Outcome}
These exchanges and the way in which \suf and \jean envision the integration of obsolete smartphones into their work is very promising and opens up many avenues for reflection. The reuse of obsolete smartphones made accessible thanks to \zombitron has clearly stimulated my two friends, and largely motivates the continuation of this work, by developing specific tools to make their two prototype projects possible. I discuss the generative potential of these discussions and beginnings of work as well as the implications for tools and future work in the next section.

\section{Discussion and future work}
In this article, I explored how a web-based approach could enable the reuse of a heterogeneous set of obsolete smartphones and tablets varying in OS and release date.
This work allowed me to explore how to create new interactive systems by combining devices of different natures as basic materials. This approach can be seen as an ``augmentation'' as described by Remy~\etal \cite{remy_addressing_2014}, as this combination is an augmentation of obsolete smartphones with other obsolete smartphones.

During my experiments on 14 obsolete smartphones, I initiated the development of a code base that can be used as the starting point for various applications.
In particular, I explored how this approach could be used in the work of musicians through discussions with \suf and \jean. This study and exchanges open up some promising perspectives on ways of pursuing the work toward a toolkit and combining it with the various software tools used by musicians, other creative practitioners and beyond.

The whole idea behind this work is to define the position of the cursor between the design opportunities and the device's capabilities.
In terms of design opportunities, this work is primarily focused on the possibilities for interaction offered by the combination of several obsolete smartphones or tablets. 
In the code base I have developed for the purposes of this study, I have integrated tactile, motion and orientation data by exploring the different constraints generated by the different ages and operating systems of smartphones.
For now, my approach has been to start with the older ones and assess what they can do, assuming that the newer ones can also do it.
Therefore, the first straightforward way of pursuing this work is to explore the other interaction possibilities offered by smartphones. For example, with the camera or microphone, which could easily be streamed using protocols such as WebRTC.
In my sample of 14 phones, many are also equipped with a brightness sensor, a fingerprint reader, an NFC chip detector, and many others which are all interaction opportunities with which to design new systems. Access to the installation of recent browsers on older devices will also make it possible to benefit from the latest APIs such as SensorAPI\footnote{\url{https://developer.mozilla.org/en-US/docs/Web/API/Sensor_APIs}}, USB\footnote{\url{https://developer.mozilla.org/en-US/docs/Web/API/WebUSB_API}} or even bluetooth\footnote{\url{https://developer.mozilla.org/en-US/docs/Web/API/Web_Bluetooth_API}}. The possibilities for interaction can also be extended by the simple fact that, by definition, any object capable of connecting to a network can be interfaced with \zombitron. It would be interesting to connect this work with the one of Norbisrath \etal~\cite{norbisrath_empowering_2025} on repurposing smartphones into gateways for IoT. More generally, exploring the limitations of a web approach based on \node is another avenue of research.

In this work, I covered the fact that the capability of devices depends on their age and OS, and I explored these aspects in my study. However, capabilities also depend on other factors, in particular their general condition. Most of the smartphones I have used have been in pretty good condition, apart from some broken screens and batteries at the end of their life. It would therefore be important to continue this work, both by continuing to explore interactive capabilities, but also by confronting devices with possibly other limitations. Further work could also look at ways in which end-users can identify and choose combinations of obsolete devices based on their performance. And extend it to other, more application-optimal techniques.

The other major part of the design opportunities is the different applications that this tool base will enable. Here, there are two different directions which are perfectly illustrated by the two distinct approaches of \jean and \suf. In one case, \jean envisages his use of \zombitron as a controller connected to \textit{Ableton}, while \suf designs \texttt{Zombee} as a standalone. On the one hand, \zombitron must be capable of interfacing with the software and, on the other, the application is executed by the smartphone. These two approaches do not involve the same things. 

In the case of \jean, this is very easy to envisage as it does not require much extra work from the smartphones. 
\zombitron is based on \texttt{websocket}, but it is also possible to communicate sensor data from other protocols such as Open Sound Control (OSC), \texttt{MQTT}, and others that are relatively easy to implement in the tool with \node. OSC is a relatively widespread protocol, not only for controlling music but also other VJing or real-time graphic composition software such as \textit{Touch Designer} or \textit{Resolume}. It would also be interesting to explore how a controller made from discarded smartphones would be considered by AV practitioners.
It would also be simple to turn it into a stream deck. 

In the case of \suf however, the underlying challenge is to explore the capabilities of smartphones and see how to optimise applications and compose them specifically for computing performance. I was confronted with latency problems in the case of the two prototypes \textit{Zombitronica} and \textit{Zombichord}, using the Web Audio library \textit{Tone.js}. At that point, the only strategy explored was to remove all the effects that consume resources. A very limiting approach in terms of functionality and creative possibilities for musicians.

This opens up avenues of exploration for future work on web approaches to performance-constrained audio synthesis on obsolete smartphones \cite{tapparo_leveraging_2023} and interactivity \cite{matuszewski_web-based_2020}.
From a web perspective in particular, one strategy to explore is the use of \textit{WebAssembly}\footnote{\url{https://webassembly.org/}}, a format that allows most hardware capabilities to be accessed from the web browser.
It is on this principle that \textit{RNBO}, the portable extension of \textit{MAX}, is based. \textit{MAX} is a popular tool among sound creators and it would therefore be a good entry point for exploring \textit{WebAssembly} on obsolete smartphones. The idea of \texttt{Zombee} represents an exciting direction to explore in terms of how a combination of more or less powerful devices can generate rich music and allow musicians a wide playing environment. 

The advantage of collaborating with practitioners is that, as well as enabling a new approach to be compared with real practices, there can be an overlap of optimisation approaches. The latter can be thought out by the tool implementation, but also by the creator themself, for example here, the idea of using granular synthesis of \suf.
The work initiated with \texttt{Zombee} is also very promising in terms of the collaborative aspect that \suf envisages for this prototype, and its approach is very much in line with the work of Golvet \etal \cite{golvet_designing_2024} and Matuszewski \etal \cite{matuszewski_interaction_2019,matuszewski_soundworks_2019,matuszewski_web-based_2020}.
It would also be interesting to combine low-tech approaches to instrument making with \zombitron, such as Nunes \etal's ingenious ideas featured in Sibilim \cite{nunes_sibilim_2019}.
Sustainability seems to have a place in NIME's thinking \cite{masu_o_2023}, and there are undoubtedly many bridges to be built with the work of this community, given their thinking and political positions on technologies \cite{jourdan_culture_2023}.
The case of music and art in general is a good research environment for this work since the communities are inclined to explore new approaches for creation. The musical axis allows many aspects at the heart of the problems of this work to be explored, namely interactivity, design and performance.

That said, this work could also have implications in other contexts, especially because of the low cost of implementation made possible by the reuse. For example, it could be used for low-cost installation in the case of documentation tools in makerspaces and laboratories~\cite{rigaud_exploring_2022}, it could also enable rapid prototyping of interfaces and be used as inexpensive teaching material. \suf and \jean both mentioned to me that it would be great to teach computer music in the workshop they or their friends are giving.

My discussions with \suf and \jean also suggest a generative power of the reuse of smartphones in creativity. Firstly, from the point of view of the aesthetics of using smartphones.
Neither \jean and \suf did particularly envisage the smartphones being concealed in an outfit, as had been explored in the case of \textit{Zombitronica} and \textit{Zombichord}. 
On the contrary, \jean explained to me how he planned to make these phones an integral part of his live show. ``\textit{It's great, I will be able to take them and put them next to each other, then put them back down again.}''. He also mentions the benefit of a smartphone as an interactive interface known by the audience, enabling them to understand what is happening on stage, as opposed to the action of turning knobs.
This relates to Mansoux \etal \cite{mansoux_permacomputing_2023}'s reflections on retro-computing aesthetics, which tend to persist in permacomputing approaches. Perhaps we could imagine an antero-computing aesthetic embodied by the smartphone or other smartwatches in the next few years. 
Another interesting generative aspect is that, although the approach is by no means a new technology, the idea of building a new system with obsolete smartphones seemed to stimulate my two friends. 
For example, gyroscopes have been around for a long time, as have smartphones connected to local networks, so \zombitron is not inventing anything that does not already exist. 
But in the case of \jean and \suf, it has generated a lot of new ideas. Perhaps the anticipation of a simple, accessible tool opens up a whole range of new possibilities. 

Although this paper is essentially the beginning of work towards a toolbox to facilitate the reuse of obsolete smartphones, and my discussions were limited to two of my musician friends, it already identifies the opportunities that a collaborative approach between researchers and designers can generate for the design of tools. This is in line with my desire to build bridges between communities, and the importance of designing tools to encourage the appropriation of sustainable techniques \cite{roedl_sustainable_2015,lu_unmaking_2024}. 

My aim is to push this study further by exploring the design of new prototypes to explore more interaction opportunities and software and hardware limitations in a research-creation process as a means of feeding this toolbox. The projects we have initiated with \suf and \jean are an embodiment of this.

\section{Conclusion}
In this article, I presented \zombitron's simple idea of making interfaces with obsolete smartphones using a high-level, web-based approach. Then, I've explored this approach in more depth, confronting it with a range of obsolete smartphones varying in OS and software version. This enabled me to start the work toward a promising toolkit to favour appropriation of the reuse of these discarded devices.
Through exchanges with two musicians leading to two speculative prototypes of musical controllers, we were able to explore how this toolkit could possibly be relevant to their practice. This work gives rise to a number of directions for research into the development of re-use techniques to encourage appropriation by creative practitioners and beyond. Firstly, it provides inspiration to pursue work on this toolkit and to continue exploring the design opportunities offered by the combination of obsolete smartphones and other obsolete devices. Secondly, it motivates the collaborative approach with artists to determine how the use of smartphones could not only enable them to adopt a more sober and less costly approach to their use of tools, but also be integrated as a central element of their performance.
More generally, this work represents a new step towards democratising the use of obsolete smartphones, despite their complexity and heterogeneity.


\begin{acks}
All the research, including the state of the art, the material and methods, the original thinking, the writing of this article and the development of the code base I used for testing, i.e. everything described outside section 3 of this article, was carried out from 28 February 2025, by me if not otherwise specified, on my own time and with my own funds.
The work on the two prototypes \textit{Zombitronica} and \textit{Zombichord} presented in section 3 was carried out while I was an employee of noesya, which retains ownership of them.
\\\\
I would like to thank Marion Jolas for her help in making the \textit{Zombichord}, Villette Makerz for hosting me as a resident during the construction of \textit{Zombitronica} and \textit{Zombichord}, La Generale for hosting me as a resident during the writing of this article and Jean Turner and s8jfou for the time they devoted to me, their interest and their help, which makes the future of this research so promising.
\end{acks}
\bibliographystyle{ACM-Reference-Format}
\bibliography{bibliography}

\newpage
\appendix
\section{Test resources}
\label{sec:supplementary}
\subsubsection*{Listing all the USB serial devices on bash}
\begin{verbatim}
ls /dev/{tty,cu}.*
\end{verbatim}

\subsubsection*{Obtaining information about an Android device from \texttt{adb}}
\begin{verbatim}
adb shell getprop ro.build.version.release
adb shell getprop ro.product.cpu.abi
\end{verbatim}

\subsubsection*{Downloading necessary packages on the Android phone via \termux}
\begin{verbatim}
pkg update --allow-unauthenticated
pkg install nodejs-lts git openssl-tool
node -v #should give the node version
npm -v #should give the npm version
\end{verbatim}

\subsubsection*{Downloading the test environment using \texttt{git}}
\begin{verbatim}
git clone --recursive [repository].git
\end{verbatim}
\texttt{[repository]} is available at \url{https://zombitron.clararigaud.com} under GPU GPL-3.0 licence.
\end{document}